\documentclass[aps,prc,preprint,showpacs,showkeys,superscriptaddress,amsmath,amssymb,floatfix,nofootinbib]{revtex4}

\usepackage[dvips]{graphicx,color}
\usepackage{dcolumn}
\usepackage{bm}

\def\pmb#1{\setbox0=\hbox{#1}%
  \kern-.025em\copy0\kern-\wd0 
  \kern.05em\copy0\kern-\wd0
  \kern-.025em\raise.0433em\box0 }

\def\lambdabar{\protect\@lambdabar}
\def\@lambdabar{%
\relax
\bgroup
\def\@tempa{\hbox{\raise.73\ht0
\hbox to0pt{\kern.25\wd0\vrule width.5\wd0
height.1pt depth.1pt\hss}\box0}}%
\mathchoice{\setbox0\hbox{$\displaystyle\lambda$}\@tempa}%
{\setbox0\hbox{$\textstyle\lambda$}\@tempa}%
{\setbox0\hbox{$\scriptstyle\lambda$}\@tempa}%
{\setbox0\hbox{$\scriptscriptstyle\lambda$}\@tempa}%
\egroup
}

\begin{document}

\preprint{J-PARC-TH-0156}

\title{\boldmath
Production Spectra of {$^3$He}($\pi$, $K$) Reactions with Continuum Discretized Coupled Channels
}

\author{Toru~Harada}%
\email{harada@osakac.ac.jp}
\affiliation{%
Osaka Electro-Communication University, Neyagawa, Osaka, 572-8530, Japan
}
\affiliation{%
J-PARC Branch, KEK Theory Center, IPNS, KEK, Tokai, Ibaraki, 319-1106, Japan.\\
}

\author{Yoshiharu~Hirabayashi}%
\affiliation{%
Information Initiative Center, 
Hokkaido University, Sapporo, 060-0811, Japan
}

\date{\today}

\begin{abstract}
We investigate theoretically $\Lambda$ production spectra of $^3{\rm He}$($\pi$,~$K$) 
reactions at $p_\pi=$ 1.05--1.20 GeV/$c$ in the distorted-wave impulse 
approximation, using the continuum-discretized coupled-channel method.
The production cross section of a $^3_\Lambda$H(1/2$^+$) ground state 
is also discussed.  

\end{abstract}
\maketitle


\section{Introduction}

It has been believed that hypertriton (${^3_\Lambda{\rm H}}$) has lifetime within a few \% of the 
free $\Lambda$ lifetime of $\tau_\Lambda=$ 263.2$\pm$2.0 ps because 
${^3_\Lambda{\rm H}}$ ($T=$ 0, $J^\pi=$ 1/2$^+$) is composed of a deuteron ($I=$ 0, $^3S_1$) 
and a $\Lambda$ hyperon loosely bound with the small $\Lambda$ binding energy 
of $B_\Lambda=$ 130$\pm$50 keV.  
It is supported by several theoretical calculations \cite{Kamada98}; see however \cite{Gal19}.
Recently, unexpected short lifetime of ${^3_\Lambda{\rm H}}$ was measured in hypernuclear production
by high-energy heavy-ion collisions \cite{HypHI13}; 
the world average lifetime of $\tau^{\rm (av)}({^3_\Lambda{\rm H}})=$ 185$^{+23}_{-28}$ ps 
is shorter than $\tau_\Lambda$ by about 30\%. 
(ALICE Collaboration \cite{ALICE18} newly reported the preliminary result of 
$\tau({^3_\Lambda{\rm H}})=$ 237$^{+33}_{-36}$ ps which is moderately closer to $\tau_\Lambda$.)
To solve the hypertriton lifetime puzzle, experimental measurements of the ${^3_\Lambda{\rm H}}$ lifetime 
are planned by $^3{\rm He}$($K^-$,~$\pi^0$) and $^3{\rm He}$($\pi^-$,~$K^0$) reactions at J-PARC. 
Moreover, HypHI Collaboration \cite{Rappold13} found a bound $nn\Lambda$ system, 
whereas theoretical calculations suggest that no $^3_\Lambda{n}$ ($T=$ 1, $J^\pi=$ 1/2$^+$) bound state 
exists \cite{Gal14}.
Therefore, it is important to investigate theoretically production of such $NN\Lambda$ systems, 
e.g., ($\pi$,~$K$) reactions on a $^3$He target, in order to settle the current problems  
related to three-body hypernuclei.

In this paper,
we focus on the $\Lambda$ production spectra of $^3{\rm He}$($\pi$,~$K$) reactions at 
$p_\pi=$ 1.05--1.20 GeV/$c$ in the distorted-wave impulse approximation (DWIA), 
using the continuum-discretized coupled-channel (CDCC) method \cite{Harada15} 
in order to well describe the $NN$ continuum states above the $N+N+\Lambda$ breakup threshold. 
We also discuss the production cross section of ${^3_\Lambda{\rm H}(1/2^+)}$ 
in $^3{\rm He}$($\pi^-$,~$K^0$) reactions, considering 
nuclear medium effects of the $\pi N \to\Lambda K$ amplitude and recoil effects.

\section{Calculations}
\label{sect:cal}

\begin{figure}[htb]
  \begin{center}
\includegraphics[width=0.8\linewidth]{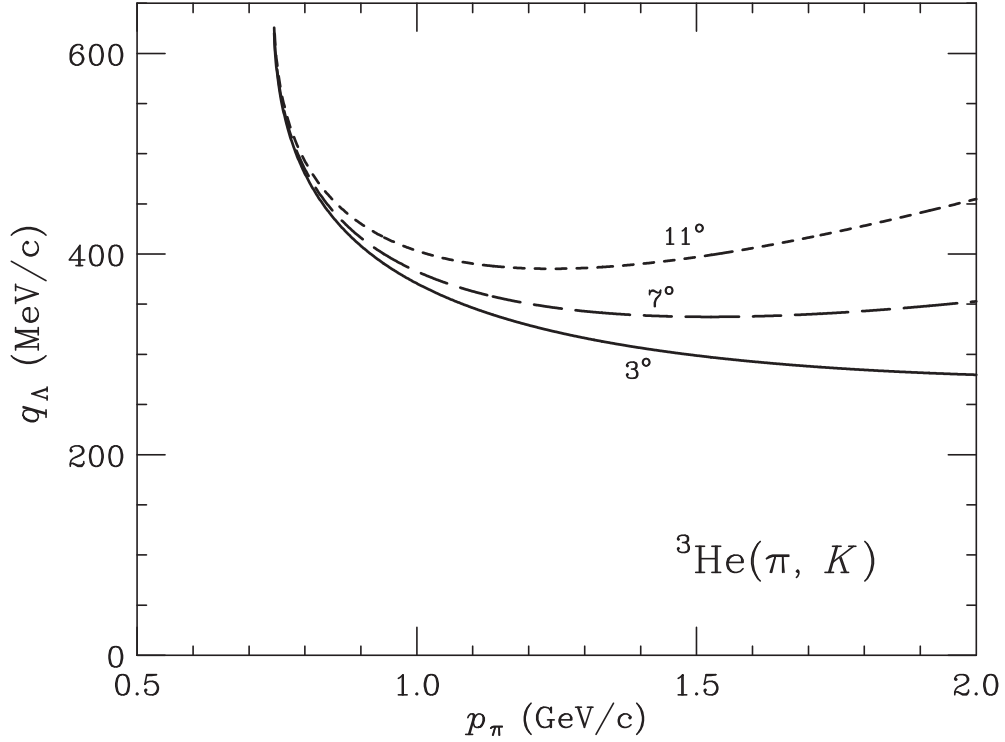}
  \end{center}
\caption{\label{fig:1}
Momentum transfer to the final state, $q_\Lambda$, for the ($\pi$,~$K$) reactions on a 
$^3$He target at scattering $K$ angles of $\theta_{\rm lab}=$ $3^{\rm o}$, $7^{\rm o}$, 
and $11^{\rm o}$ in the laboratory frame, 
as a function of the incident pion momentum $p_{\pi}$. 
  }
\end{figure}

\begin{figure}[htb]
  \begin{center}
\includegraphics[width=0.8\linewidth]{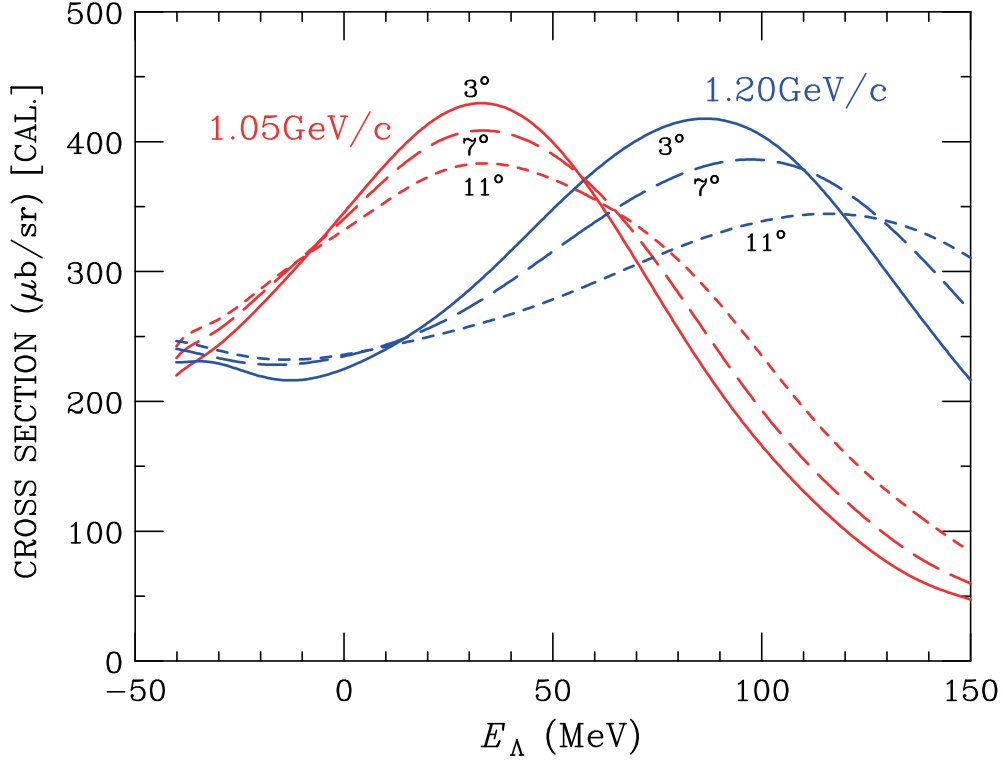}
  \end{center}
\caption{\label{fig:2}
Calculated laboratory cross sections for the $\pi N\to\Lambda K$ reactions in nuclear medium 
at $p_{\pi}=$ 1.05 and 1.20 GeV/$c$ with $\theta_{\rm lab}=$ $3^{\rm o}$, $7^{\rm o}$, 
and $11^{\rm o}$, as a function of $E_\Lambda$. 
The optimal Fermi averaging \cite{Harada04} is used.
}
\end{figure}

Inclusive differential cross sections for nuclear ($\pi$,~$K$) reactions in the laboratory frame 
within the DWIA \cite{Harada15} are given by (in units $\hbar=c=1$)
\begin{eqnarray}
   {{d}^2{\sigma} \over {d}E_K {d}\Omega_K}
= \beta {1 \over {[J_A]}}
  \sum_{M_A}\sum_B \vert\langle {\Psi}_B \vert
  \hat{F} \vert {\Psi}_A \rangle\vert^{2} \delta (E_{K}+E_{B}-E_{\pi}-E_{A}), 
\label{eqn:e1}
\end{eqnarray}
where $[J]=2J+1$, $\beta$ is a kinematical factor, and $E_{K}$, $E_{\pi}$, $E_{B}$ and $E_{A}$ are 
energies of outgoing $K$, incoming $\pi$, hypernuclear states and the target nucleus, respectively; 
$\Psi_B$ and $\Psi_A$ are wavefunctions of hypernuclear states and the target nucleus, respectively.
$\hat{F}$ is a strangeness-exchange external operator given by
\begin{eqnarray}
\hat{F}
 = \int d{\bm r} \> \chi_{K}^{(-) \ast}({\bm p}_K,{\bm r})
\chi_{\pi}^{(+)}({\bm p}_\pi,{\bm r}) \sum_{j=1}^A \overline{f}_{\pi N\to\Lambda K}
\delta ({\bm r}-{\bm r}_{j}){\hat O}_j, 
\label{eqn:e2}
\end{eqnarray}
where $\chi_{K}^{(-) \ast}$ and $\chi_{\pi}^{(+)}$ are distorted waves for 
outgoing $K$ and incoming $\pi$, respectively, which are calculated 
with the help of the eikonal approximation. 
${\hat O}_j$ is a baryon operator changing $j$th nucleon into a $\Lambda$ hyperon in the nucleus.
Figure \ref{fig:1} displays the momentum transfer to the final state, 
$q_\Lambda =|{\bm p}_{\pi}-{\bm p}_{K}|$, as a function of 
the incident pion momentum $p_{\pi}$, where ${\bm p}_{\pi}$ and ${\bm p}_{K}$ are 
the laboratory momenta of $\pi$ and $K$ in the nuclear reaction, respectively.
$\overline{f}_{\pi N\to\Lambda K}$ is the $\pi N\to\Lambda K$ amplitude 
in nuclear medium, which is obtained by the optimal Fermi-averaging method \cite{Harada04}.
It should be noticed that strong $E_\Lambda$ dependence appears 
in $\overline{f}_{\pi N\to\Lambda K}$ 
because the elementary cross sections for the $\pi N\to\Lambda K$ reactions 
depend on the incident pion momentum \cite{Sotona89}; 
we confirm that the optimal Fermi-averaged cross sections of 
${{d}{\sigma}/{d}\Omega}= \beta |\overline{f}_{\pi N\to\Lambda K}|^2$ 
at $p_{\pi}=$ 1.05 and 1.20 GeV/$c$ have strong $E_\Lambda$ dependence \cite{Harada04}, 
as shown in Fig.~\ref{fig:2}.

\begin{figure}[t]
  \begin{center}
\includegraphics[width=0.8\linewidth]{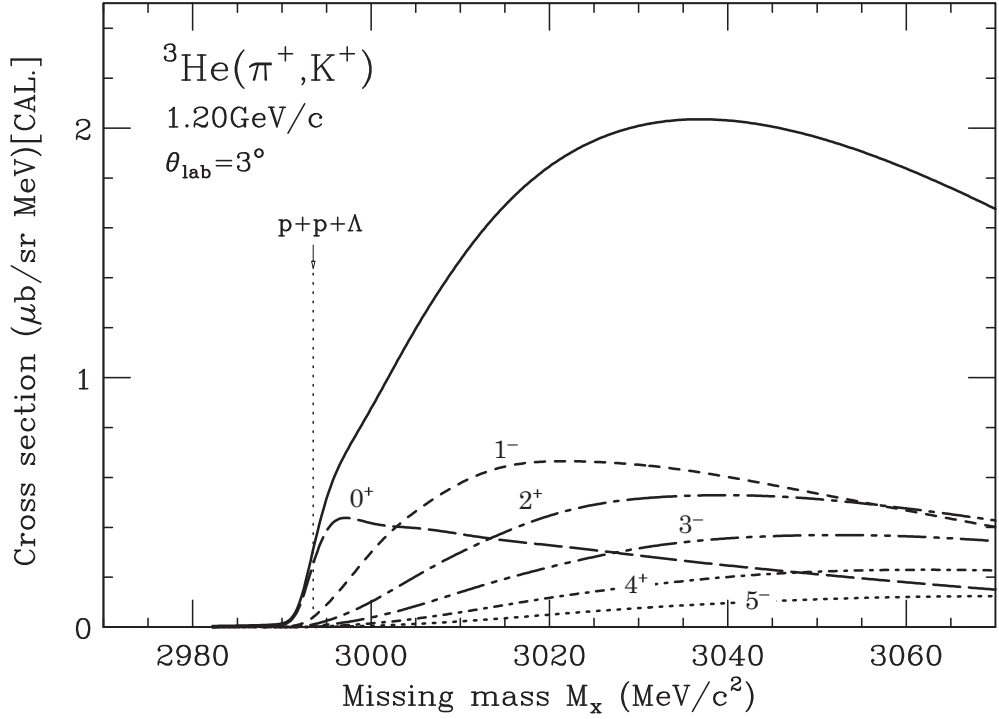}
  \end{center}
\caption{\label{fig:3}
Calculated inclusive $K^+$ spectrum in the $^3$He ($\pi^+$,~$K^+$) reaction
at 1.20 GeV/c, $\theta_{\rm lab}=$ 3$^\circ$, 
together with partial-wave components $L^\pi$ and $S=1/2$, 
as a function of the missing mass $M_{\rm x}$.
The spectrum is taken into account a detector resolution of 4 MeV FWHM.
}
\end{figure}

Hypernuclear final states are considered as three-body $NN\Lambda$ systems 
in the $^3$He($\pi$,~$K$) reactions, involving continuum states above the $N+N+\Lambda$ threshold. 
Here we employ the CDCC method \cite{Kamimura86} in order to well describe the 
$NN$ continuum states as breakup channels.
Thus the wavefunctions ${\Psi}_B$ with $J^\pi$ in the $LS$-coupling scheme can be written as
\begin{eqnarray}
&&{\Psi}_B 
 \simeq  {\Psi}^{\rm CDCC}_B({\bm r},{\bm R})
 =  \sum_{\alpha=1}^{N_{\rm max}}\sum_{\ell_2=0}^{\ell_{\rm max}} \left[ 
   \bigl[\tilde{\phi}_{\alpha,\ell_2}^{(2N)}({\bm r})\otimes
   \varphi^{(\Lambda)}_{\alpha,\ell_\Lambda}
   ({\bm R}) \bigl]_{L_B} \otimes X^B_{I_\alpha,S_\alpha}
    \right]_{J_B}^{M_B}, 
\label{eqn:e3}
\end{eqnarray}
where 
$\tilde{\phi}_{\alpha,\ell_2}^{(2N)}({\bm r})$ is the $NN$ wavefunction having 
bound and continuum-discretized states for angular momentum $\ell_2$, 
spin $^1S_0$ or $^3S_1$ in channel $\alpha$, 
$\varphi^{(\Lambda)}_{\alpha,\ell_\Lambda}({\bm R})$ is the relative wavefunction 
between $NN$ and $\Lambda$ with $\ell_\Lambda$, 
and $X^B_{I_\alpha,S_\alpha}$ is the isospin-spin function for $NN\Lambda$.
Because we omit spin-flip processes in the ($\pi$,~$K$) reactions on the $^3$He (1/2$^+$) target, 
the final states on $J^\pi=|L^\pi \pm 1/2|$ with $L^\pi=$ 0$^+$, 1$^-$, 2$^+$, $\cdots$, 
and $S=$ 1/2 can be populated.
We obtain $\varphi^{(\Lambda)}_{\alpha,\ell_\Lambda}$, solving a coupled-channel equation 
with the potential $U_{\alpha\alpha'}$ given by the microscopic 
$2N$-$\Lambda$ folding model; 
\begin{equation}
U_{\alpha\alpha'}(R)=
\int  \!\rho_{\alpha\alpha'}({\bm r}) 
     \Bigl(\overline{v}_{\Lambda N}({\bm R}-{\bm r}/2)
     +\overline{v}_{\Lambda N}({\bm R}+{\bm r}/2)\Bigr)
     d{\bm r},
\label{eqn:e4}
\end{equation}
where $\rho_{\alpha\alpha'}({\bm r})$ is the nucleon or transition density, 
and $\overline{v}_{\Lambda N}$ is the spin-averaged $\Lambda N$ potential. 
For the $\Lambda N$ potential $v_{\Lambda N}$, we assume a single Gaussian form 
which reproduces the scattering length and the effective range in $\Lambda p$ scattering
at low energies, fitting into those of NSC97f. 
We can also reproduce the experimental value of $B_\Lambda=$ 0.13 MeV for $^3_\Lambda$H(1/2$^+$) 
when we slightly modify the strengths of $v_{\Lambda N}$ by a factor of 0.92.
We rewrite a sum over the final states in Eq.~(\ref{eqn:e1}) as
\begin{equation}
 \sum_{B}  \vert {\Psi}_B \rangle \langle {\Psi}_B \vert \delta (E_{B}-E_{A}-\omega)
  = -{1 \over \pi} {\rm Im}\,\hat{G}(\omega), 
\label{eqn:e5}
\end{equation}
where $\hat{G}(\omega)$ is a complete Green's function for the $2N$-$\Lambda$ systems 
given by CDCC wavefunctions.
Therefore, the inclusive differential cross sections are obtained by the Green's function 
method \cite{Morimatsu85}.

\begin{figure}[bht]
  \vspace{5mm}
  \begin{center}
\includegraphics[width=0.8\linewidth]{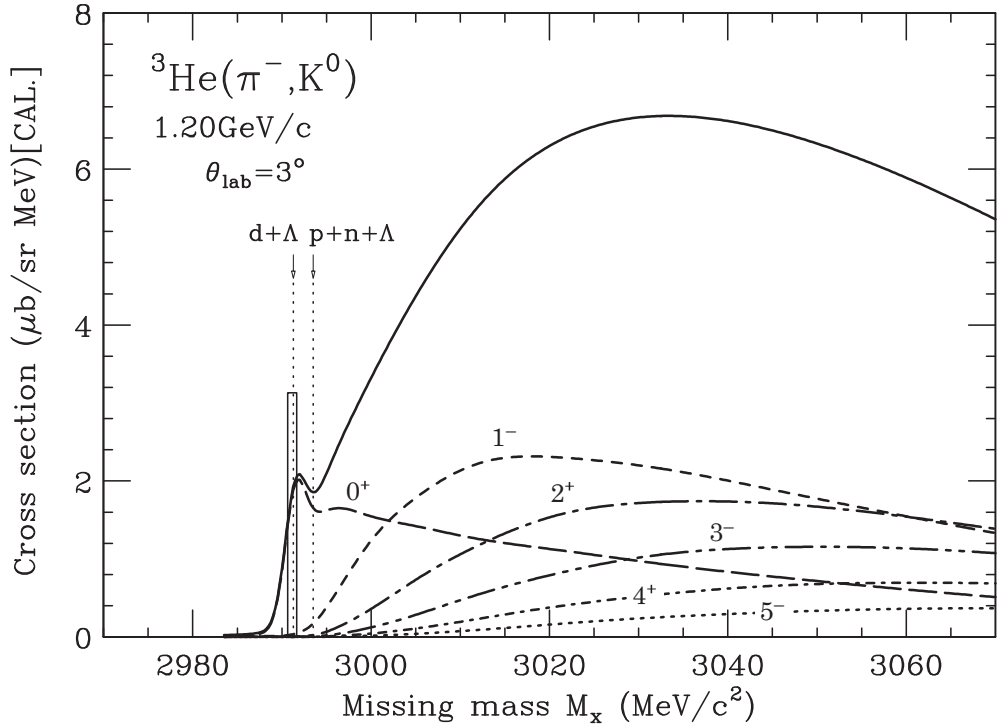}
  \end{center}
  \caption{\label{fig:4}
Calculated inclusive $K^0$ spectrum in the $^3$He ($\pi^-$,~$K^0$) reaction 
at 1.20 GeV/c, $\theta_{\rm lab}=$ 3$^\circ$. 
The bin with a finite width of 1 MeV at $M_{\rm x}=$ 2991.2 MeV/$c^2$  
denotes the integrated cross section of $^3_\Lambda$H(1/2$^+$). 
See also the caption of Fig.~\ref{fig:3}.
  }
\end{figure}

\section{Results and Discussion}

\begin{table*}[bht]
\caption{
\label{tab:1}
Calculated results of the integrated production cross sections of the 
${^3_\Lambda{\rm H}(1/2^+)}$ ground state, $d\sigma/d\Omega$\,(${^3_\Lambda{\rm H}}$), 
in the $^3$He($\pi^-$, $K^0$) reactions at $p_{\pi}=$ 1.05 and 1.20 GeV/$c$. 
}
\begin{ruledtabular}
\begin{tabular}{ccccccccccc}
$p_{\pi}$ & $\theta_{\rm lab}$ 
&& \multicolumn{2}{c}{$(d\sigma/d\Omega)^{{\small \pi N\to\Lambda K}}_{\rm elem.}$\, ($\mu$b/sr)}
&& $q_\Lambda$   &  $q_\Lambda^{\rm eff}$\tablenotemark[3]
&& \multicolumn{2}{c}{$d\sigma/d\Omega$\,(${^3_\Lambda{\rm H}}$)\, ($\mu$b/sr)} \\
\noalign{\smallskip}
 \cline{4-5}  \cline{10-11}  
\noalign{\smallskip}
(GeV/$c$)  & (degree)  
&& free\tablenotemark[1] & in medium\tablenotemark[2]
&& (MeV/$c$)  & (MeV/$c$) 
&& w/o recoil & full \\
\noalign{\smallskip}\hline\noalign{\smallskip}
1.05  &  3 &&  463  &  344  &&  354 &  236  && 0.15  & 3.07 \\
      &  7 &&  459  &  340  &&  369 &  246  && 0.10  & 2.42 \\
      & 11 &&  450  &  332  &&  393 &  262  && 0.05  & 1.59 \\
1.20  &  3 &&  292  &  225  &&  326 &  217  && 0.22  & 3.13 \\
      &  7 &&  287  &  235  &&  350 &  233  && 0.13  & 2.24 \\
      & 11 &&  277  &  236  &&  383 &  255  && 0.05  & 1.33 \\
\end{tabular}
\end{ruledtabular}
\tablenotetext[1]{$\beta|{f}_{\pi^-p\to\Lambda K^0}|^2$ is used. }
\tablenotetext[2]{$\beta|\overline{f}_{\pi^-p\to\Lambda K^0}|^2$ is used.}
\tablenotetext[3]{$q_\Lambda^{\rm eff}\approx (M_{\rm C}/M_{\rm A})q_\Lambda$ where a recoil factor of $M_{\rm C}/M_{\rm A}$ is equal to 2/3 for the $^3$He target.}           
\end{table*}

In the $^3$He($\pi^+$,~$K^+$) reactions, we consider the production of the $pp\Lambda$ states 
with only $NN$, $^1S_0$ components.
Figure \ref{fig:3} displays the calculated inclusive $K^+$ spectrum of the $^3$He($\pi^+$,~$K^+$) 
reaction at $p_{\pi}=$ 1.20 GeV/$c$, $\theta_{\rm lab}=$ 3$^{\rm o}$, 
together with partial-wave components of the spectrum. 
Considering large momentum transfer of $q_\Lambda \simeq$ 360 MeV/$c$ in exothermic ($\pi$,~$K$) reactions,
we find that many partial waves moderately contribute to the spectrum; 
there appears an enhancement of the $T=$ 1, $J^\pi=$ 1/2$^+$ ($L^\pi=$ 0$^+$, $S=$ 1/2) component just above 
the $p+p+\Lambda$ threshold.
This enhancement may indicate that a pole of the $s$-wave $pp\Lambda$ resonance or virtual state 
resides near the $p+p+\Lambda$ threshold, as suggested by several three-body calculations. 

In the $^3$He($\pi^-$,~$K^0$) reactions, we study the production of the $pn\Lambda$ states 
with $NN$, $^3S_1$ and $^1S_0$ components, 
which include the $^3_\Lambda{\rm H}$\,($T=$ 0, $J^\pi=$ 1/2$^+$) ground state. 
Figure~\ref{fig:4} displays the calculated inclusive $K^0$ spectrum of the $^3$He($\pi^-$,~$K^0$) 
reaction at $p_{\pi}=$ 1.20 GeV/$c$, $\theta_{\rm lab}=$ 3$^{\rm o}$, together with partial-wave 
components of the spectrum. We find that the integrated production cross section 
of $^3_\Lambda{\rm H}$ amounts to $d\sigma/d\Omega\,(^3_\Lambda{\rm H})=$ 3.13 $\mu$b/sr.
In Table \ref{tab:1}, we show the values of $d\sigma/d\Omega\,(^3_\Lambda{\rm H})$ 
at $p_{\pi}=$ 1.05 and 1.20 GeV/$c$ with $\theta_{\rm lab}=$ 3$^{\rm o}$, 7$^{\rm o}$, and 11$^{\rm o}$
in order to see the sensitivity to the in-medium $\pi N\to\Lambda K$ amplitudes 
and to momentum transfers.
We find that the values of $d\sigma/d\Omega\,(^3_\Lambda{\rm H})$ at 1.05 and 1.20 GeV/$c$ are similar, 
although the values of $(d\sigma/d\Omega)^{{\small \pi N\to\Lambda K}}_{\rm elem.}$ 
near the $\Lambda$ threshold at 1.05 GeV/$c$ are about 1.5 times as large as those at 1.20 GeV/$c$. 
This is caused by the fact that the former is larger than the latter in terms of $q^{\rm eff}_\Lambda$.
If the recoil effects are switched off ($M_C/M_A = 2/3$ is replaced by 1),
the values of $d\sigma/d\Omega\,(^3_\Lambda{\rm H})$ are reduced by an order of magnitude or more.
Hence we recognize that the recoil effects for production on the light target 
as $^3{\rm He}$ are important in the nuclear ($\pi$,~$K$) reactions.

\section{Summary}

We have shown the calculated $\Lambda$ production spectra of the $NN\Lambda$ systems  
in the $^3{\rm He}$($\pi$,~$K$) reactions at 1.05--1.20 GeV/$c$ 
with CDCC which describes the $NN$ continuum states above the $N+N+\Lambda$ 
breakup threshold. 
The production cross section of ${^3_\Lambda{\rm H}(1/2^+)}$ in the ($\pi^-$,~$K^0$) reaction
is evaluated, e.g., $d\sigma/d\Omega\,({^3_\Lambda{\rm H}})\simeq$ 3 $\mu$b/sr at 1.05--1.20 GeV/$c$, 
$\theta_{\rm lab}=$ 3$^{\rm o}$. 
The recoil effects are very important to the production with the light nuclear 
target as $^3{\rm He}$, as well as the medium effects of the $\pi N\to\Lambda K$ amplitudes 
for nuclear ($\pi$,~$K$) reactions.
More precise analysis on convergence of the CDCC model space depending on 
($k_{\rm max}$, $\ell_{\rm max}$) should be needed. 
This investigation is in progress.

The authors would like to thank H.~Tamura and A.~Gal for valuable comments.
This work was supported by JSPS KAKENHI Grant Numbers JP16K05363.

\clearpage

\end{document}